\documentclass{jpp}

\usepackage{graphicx}
\usepackage{caption}
\usepackage{subcaption}
\usepackage{dcolumn}
\usepackage{bm}
\usepackage[dvipsnames]{xcolor}
\usepackage{amssymb}
\usepackage{amsmath}
\usepackage{verbatim}
\usepackage{appendix}
\usepackage{listings}
\usepackage{units}
\usepackage{csquotes}
\usepackage{tikz}

\usepackage{hyperref}

\usepackage{etoolbox}
\makeatletter
\patchcmd{\hyper@makecurrent}{%
    \ifx\Hy@param\Hy@chapterstring
        \let\Hy@param\Hy@chapapp
    \fi
}{%
    \iftoggle{inappendix}{
        \@checkappendixparam{chapter}%
        \@checkappendixparam{section}%
        \@checkappendixparam{subsection}%
        \@checkappendixparam{subsubsection}%
        \@checkappendixparam{paragraph}%
        \@checkappendixparam{subparagraph}%
    }{}%
}{}{\errmessage{failed to patch}}

\newcommand*{\@checkappendixparam}[1]{%
    \def\@checkappendixparamtmp{#1}%
    \ifx\Hy@param\@checkappendixparamtmp
        \let\Hy@param\Hy@appendixstring
    \fi
}
\makeatletter

\newtoggle{inappendix}
\togglefalse{inappendix}

\apptocmd{\appendix}{\toggletrue{inappendix}}{}{\errmessage{failed to patch}}
\apptocmd{\subappendices}{\toggletrue{inappendix}}{}{\errmessage{failed to patch}}

\newcommand*\obar[2][0.75]{
    \sbox{\myboxA}{$\m@th#2$}%
    \setbox\myboxB\null
    \ht\myboxB=\ht\myboxA%
    \dp\myboxB=\dp\myboxA%
    \wd\myboxB=#1\wd\myboxA
    \sbox\myboxB{$\m@th\overline{\copy\myboxB}$}
    \setlength\mylenA{\the\wd\myboxA}
    \addtolength\mylenA{-\the\wd\myboxB}%
    \ifdim\wd\myboxB<\wd\myboxA%
       \rlap{\hskip 0.5\mylenA\usebox\myboxB}{\usebox\myboxA}%
    \else
        \hskip -0.5\mylenA\rlap{\usebox\myboxA}{\hskip 0.5\mylenA\usebox\myboxB}%
    \fi}
\makeatother

\definecolor{colorexample}{RGB}{250,143,56}

\definecolor{ecolor}{HTML}{4477AA}
\definecolor{Hcolor}{HTML}{117733}
\definecolor{Necolor}{HTML}{88CCEE}
\definecolor{Hecolor}{HTML}{DDCC77}
\definecolor{Ccolor}{HTML}{CC6677}

\renewcommand{\vec}[1]{\boldsymbol{#1}}

\newcommand{\e}{\ensuremath{\mathrm{e}}}

\renewcommand{\d}{\ensuremath{\mathrm{d}}}

\newcommand{\remove}[1]{{}}

\newcommand{\lang}{\left\langle}
  \newcommand{\rang}{\right\rangle}

\usepackage{stackengine}
\stackMath

\newcommand{\appref}[1]{\hyperref[#1]{Appendix~\ref{#1}}}

\newcommand{\tikzdashed}[1]{\protect\tikz[baseline=-.5ex,ultra thick]{\protect\draw[line width=1pt,dash pattern=on 2pt off 1.5pt, color=#1] (0,0) -- (0.35,0);}}
\newcommand{\tikzsolid}[1]{\protect\tikz[baseline=-.5ex,ultra thick]{\protect\draw[line width=1pt, color=#1] (0,0) -- (0.35,0);}}

\usepackage{morefloats}

\title{The importance of the classical channel in the impurity transport of optimized stellarators}
\author{S.~Buller, A.~Moll\'{e}n, S.L.~Newton, H.M.~Smith, I.~Pusztai}

\begin{document}
\maketitle
\begin{abstract}
  In toroidal magnetic confinement devices, such as tokamaks and stellarators, neoclassical transport is usually an order of magnitude larger than its classical counterpart.
  However, when a high-collisionality species is present in a stellarator optimized for low Pfirsch-Schl\"uter current, its classical transport can be comparable to the neoclassical transport.
  In this letter, we compare neoclassical and classical fluxes and transport coefficients calculated for Wendelstein 7-X (W7-X) and Large Helical Device (LHD) cases. 
  In W7-X, we find that the classical transport of a collisional impurity is comparable to the neoclassical transport for all radii, while it is negligible in the LHD cases, except in the vicinity of radii where the neoclassical transport changes sign.
  In the LHD case, electrostatic potential variations on the flux-surface significantly enhance the neoclassical impurity transport, while the classical transport is largely insensitive to this effect in the cases studied.
\end{abstract}

\section{Introduction}

The most developed concepts for achieving controlled thermonuclear fusion are the tokamak and stellarator.
Both the tokamak and the stellarator utilize a strong toroidal magnetic field to confine a hot plasma in which fusion reactions take place.

When such a plasma is in a steady-state, loss of particles and energy mainly occurs as a result of micro-turbulence, collisions, or direct losses of particles on unconfined orbits. 
The two latter processes -- and the resulting transport of particles and heat -- is referred to as \emph{collisional transport}, and can be modeled within the framework of drift-kinetics. Historically this is the dominant transport channel in the core of stellarators because of the large transport due to particles on unconfined orbits \citep{fp2012iaea}.


Collisional transport can be further separated into two additive components: classical transport, which is due to the gyro-motion of particles around the magnetic field-lines, and neoclassical transport, which is due to the complex orbits carried out by the center of gyration as it moves in the magnetic field. The latter typically leads to much larger transport than the former \citep{ps1962}, and also accounts for the unconfined orbits in stellarators, with a very strong unfavorable scaling towards reactor-relevant high temperatures. Thus, much effort has been devoted to reducing the neoclassical transport in stellarators, resulting in optimized stellarators such as Wendelstein 7-X (W7-X) \citep{nuhrenberg1986}, while classical transport is often neglected.

However, it has not been widely appreciated that, as a result of optimizing for low neoclassical transport and Pfirsch-Schl\"{u}ter current, the neoclassical transport of impurities in W7-X can now be comparable to the often neglected classical transport. The main purpose of the present note is to raise attention to this circumstance.

To understand why the classical transport is relevant in an optimized stellarator, we employ recent analytical results on neoclassical transport for a collisional impurity \citep{braun2010a,helanderPRL2017,newton2017} to show that the ratio of classical to neoclassical fluxes is proportional to a geometrical factor \eqref{eq:nctoc}, which turns out to be larger than one in W7-X.

Motivated by these results, we present a general expression for the classical transport, using the linearized Fokker-Planck operator and allowing for an arbitrary number of species. 
The employed collision operator is frequently used in modern neoclassical solvers, and the results can thus be directly compared with the output from such codes. In the final sections, we look at a few example magnetic configurations, where we compare the magnitude of the classical transport to that of the neoclassical transport calculated with the \textsc{Sfincs}\footnote{Available at: \url{https://github.com/landreman/sfincs} (verified 2019-01-28)} drift-kinetic solver \citep{sfincs2014}, and investigate the collisionality dependence of the ratio of classical to neoclassical transport. 

\section{Motivation}
\label{sec:motivation}
Before performing a detailed analysis, it is useful to consider a simple (but experimentally relevant) limit, where the importance of classical transport in a stellarator is apparent. For this purpose, we summarize results from earlier work \citep{buller2018jpp,braun2010a,helanderPRL2017}.

At fusion-relevant temperatures, the bulk hydrogen species of the confined plasma will be in a low-collisionality regime. However, as the collisionality increases with charge,
high-$Z$ impurities (with $Z$ being the charge number) can still have high collisionality. 
Such impurities can occur, for example, in experiments using tungsten plasma-facing components, which is the favoured material for the divertor of future fusion reactors \citep{bolt2002}.
These plasmas will thus be in a \emph{mixed-collisionality regime}, with low-collisionality bulk and high-collisionality impurity ions.

In this regime, the ratio of classical to neoclassical impurity particle fluxes calculated from the mass-ratio expanded collision operator is given by a purely geometrical factor \citep{buller2018jpp}
\begin{equation}
  \frac{\lang \vec{\Gamma}_z \cdot \nabla \psi \rang^{\text{C}}}{\lang \vec{\Gamma}_z \cdot \nabla \psi \rang^{\text{NC}}} = \frac{\lang j_\perp^2\rang \lang B^2 \rang}{\lang j_\|^2 \rang \lang B^2 \rang - \lang j_\| B \rang^2}. \label{eq:nctoc}
\end{equation}
Here, $\psi$ is a radial coordinate (a flux surface label), $\lang \cdot \rang$ is the flux-surface average, $\vec{\Gamma}_z$ is the flux of impurity ions, $\langle \vec{\Gamma}_z \cdot \nabla \psi \rangle^{\text{(N)C}}$ is the radial (neo)classical impurity flux averaged over the flux-surface, $\vec{B}$ is the magnetic field, $B=|\vec{B}|$, and $\vec{j}$ is the current density, here defined by $\vec{j} \times \vec{B}= \nabla p(\psi)$, $\nabla \cdot \vec{j}=0$; with $j_\|$ and $j_\perp$ being the current components parallel and perpendicular to $\vec{B}$, and $p$ the total pressure.

\autoref{eq:nctoc} also enters into the ratio of classical and neoclassical transport at yet higher collisionalities: in the Pfirsch-Schl\"{u}ter regime, where both bulk and impurity ions are collisional. This can be shown using the expression for neoclassical transport derived by \citet{braun2010a} together with the expression for classical transport in, for example in \citet{buller2018jpp}.
For stellarators optimized for low $j_\|/j_\perp$ (such as W7-X), the \eqref{eq:nctoc} ratio will be large and classical transport will thus dominate at high collisionality. This will be verified by numerical simulations in \autoref{sec:colscan}.

\section{Linearized Fokker-Planck operator}
In this section, we write down the classical particle and heat transport due to a linearized Fokker-Planck operator. The flux-surface averaged radial classical transport of particles and energy is given by
\begin{align}
  \Gamma_a^{\text{C}}  \equiv \lang \vec{\Gamma}_a \cdot \nabla \psi \rang^{\text{C}} \equiv& \lang \frac{\vec{b} \times \nabla \psi}{Z_a e B} \cdot\vec{R}_a \rang,\label{eq:Gamma} \\
  Q_a^{\text{C}}  \equiv \lang \vec{Q}_a \cdot \nabla \psi \rang^{\text{C}} \equiv& \lang \frac{\vec{b} \times \nabla \psi}{Z_a e B} \cdot\vec{G}_a \rang,\label{eq:Q}
\end{align}
where we have introduced the \emph{friction force} and \emph{energy-weighted friction force}
\begin{align}
  \vec{R}_a &\equiv \int  m_a \vec{v} C[f_{a}]\d^3 v, \label{eq:R}\\
  \vec{G}_a &\equiv \int \frac{m_a v^2}{2}  m_a \vec{v} C[f_{a}]\d^3 v \label{eq:G}.
\end{align}
Here, $C[f_a] = \sum_b C_{ab}[f_a,f_b]$ is the Fokker-Planck collision operator, accounting for the collisions of all species '$b$' with species '$a$'; $f_{a}$ the distribution function of species '$a$', with mass $m_a$ and charge $Z_a e$, with $e$ the elementary charge; the integral is over all velocities $\vec{v}$. In a confined plasma, the distribution functions are close to a Maxwell-Boltzmann distribution $f_{a0}$, such that $f_a = f_{a0} + f_{a1}$, and $f_{a1}$ satisfies $f_{a1}/f_{a0} \ll 1$. For later reference, we also define the classical conductive heat flux $q_a^{\text{C}} = Q_a^{\text{C}} - \frac{5}{2} T_a \Gamma_a^{\text{C}}$, where $T_a$ is the temperature of species '$a$'.

For a magnetized plasma, it is useful to separate out the dependence of the distribution function on the gyrophase. Only the gyrophase-dependent part of $f$, which we denote by $\tilde{f}$, contributes to $\vec{R}$ and $\vec{G}$ perpendicular to the magnetic field, and thus to the classical fluxes \eqref{eq:Gamma}--\eqref{eq:Q}. For a magnetized plasma with an isotropic Maxwellian, it is well-known that \citep{hazeltine1973}
\begin{equation}
\tilde{f}_{a1} = -\vec{\rho}_a \cdot \nabla f_{a0}, \label{eq:ftilde}
\end{equation}
where $\vec{\rho}_a = \vec{B} \times \vec{v} m_a/(Z_aeB^2)$ is the gyro-radius vector.

With \eqref{eq:ftilde}, we can readily evaluate the classical transport given by \eqref{eq:Gamma}--\eqref{eq:G}. Lately in stellarator research, the importance of flux-surface variation of the electrostatic potential has been recognized \citep{garcia2017}; such effects can be incorporated into the classical transport by including the flux-surface varying part of the potential in the Maxwell-Boltzmann distribution $f_0$ \citep{hintonWong1985}
\begin{equation}
  f_{0} = \eta(\psi) \left(\frac{m}{2\pi T} \right)^{3/2} \exp{\left(-\frac{m v^2}{2T} - \frac{Z e\tilde{\Phi} }{T}\right)},
  \label{eq:MB}
\end{equation}
where $\Phi$ is the electrostatic potential, $\tilde{\Phi} = \Phi-\lang \Phi \rang$, and we have introduced the \emph{pseudo-density}
\begin{equation}
    \begin{aligned}
    \eta(\psi) &\equiv n \e^{\frac{Ze\tilde{\Phi}}{T}},
  \end{aligned}\label{eq:eta}
\end{equation}
with $n$ the density.
In terms of gradients of $\eta$, $T$ and $\Phi$, the gradient in \eqref{eq:ftilde} thus becomes, 
\begin{equation}
 \nabla f_0 = \nabla \psi \frac{\p f_{0}}{\p \psi} = \nabla \psi f_0\left[\frac{\d \ln\eta}{\d \psi} + \frac{Z_a e}{T_a} \frac{\p \tilde{\Phi}}{\p \psi}  + \frac{Z_a e \tilde{\Phi}}{T_a} \frac{\d \ln T_a}{\d \psi}
 + \left(\frac{m_a v^2}{2T_a} - \frac{3}{2}\right)  \frac{\d \ln T_a}{\d \psi}\right]. \label{eq:dpsif0}
\end{equation}

With this $\nabla f_0$, the resulting classical fluxes can be calculated using Braginskii matrices (as in, for example, \citet{newton2006}), resulting in
\begin{align}
 \Gamma_a^{\text{C}}  = & \frac{m_a}{Z_a e^2}\sum_b \frac{1}{\tau_{ab} n_b}\left\{\lang  n_a n_b\frac{ |\nabla \psi|^2}{B^2} \rang \right.\left.M_{ab}^{00} \left(\frac{ T_a}{Z_a}  \frac{\d \ln\eta_a}{\d \psi} -\frac{ T_b}{Z_b}  \frac{\d \ln\eta_b}{\d \psi}\right) \right.\nonumber \\
    + & \left.  \lang n_a n_b\frac{  |\nabla \psi|^2}{B^2} e\tilde{\Phi} \rang \right.\left.M_{ab}^{00} \left(\frac{\d \ln T_a}{\d \psi} -\frac{\d \ln T_b}{\d \psi}\right) \right.\label{eq:fluxgS} \\
    + & \left. \lang n_a n_b \frac{|\nabla \psi|^2 }{B^2}\rang\right.\left.\left[  \left(M_{ab}^{00} - M_{ab}^{01}\right)\frac{ T_a}{Z_a}\frac{\d \ln T_a}{\d \psi} \right.  - \left. \left(M_{ab}^{00} - \frac{m_a T_b}{m_b T_a} M_{ab}^{01}\right)\frac{T_b}{Z_b}\frac{\d \ln T_b}{\d \psi}\right]\right\},\nonumber \\
  q_a^{\text{C}}  = &  -\frac{T_a m_a}{Z_a e^2} \sum_b \frac{1}{\tau_{ab} n_b} \left\{
    \lang n_a n_b\frac{|\nabla \psi|^2}{B^2 } \rang \right. \left.M_{ab}^{01}\left(\frac{T_a}{Z_a}  \frac{\d \ln\eta_a}{\d \psi} - \frac{T_b}{Z_b}\frac{\d \ln\eta_b}{\d \psi}\right) \right. \nonumber \\
    + &  \left.\lang n_a n_b\frac{|\nabla \psi|^2}{B^2 } e \tilde{\Phi} \rang \right. \left.M_{ab}^{01} \left(\frac{\d \ln T_a}{\d \psi}-\frac{\d \ln T_b}{\d \psi}\right) \right. \label{eq:fluxqS} \\
    + & \left. \lang n_a n_b\frac{|\nabla \psi|^2}{B^2 } \rang \right. \left.\left[\left(M_{ab}^{01} - M_{ab}^{11}\right)\frac{ T_a}{Z_a} \frac{\d \ln T_a}{\d \psi}\right.
     \left.- \left(M_{ab}^{01} + N_{ab}^{11}\right)\frac{T_b}{Z_b}\frac{\d \ln T_b}{\d \psi}\right] \right\}, \nonumber
\end{align}
where $M_{ab}^{jk}$ are the Braginskii matrix elements \citep{braginskii1958}, defined in Appendix \ref{sec:bme}, using the same notation as \citet{helander2005}; the collision time $\tau_{ab}$ is defined as
\begin{equation}
  \frac{1}{\tau_{ab} n_b} \equiv \frac{\sqrt{2}Z_a^2 Z_b^2 e^4 \ln \Lambda}{12 \pi^{3/2} \epsilon_0^2 m_a^{1/2} T_a^{3/2}},
\end{equation}
where $\ln \Lambda$ the Coulomb logarithm, and $\epsilon_0$ the vacuum permittivity. These expressions are valid for all collisionalities. 
In \eqref{eq:fluxgS} and \eqref{eq:fluxqS}, the
effect of $\tilde{\Phi}$ is to induce a weighting over the flux-surface due to the flux-surface variation of $n_a$ and its radial gradient. Note that the radial electric field (from $\lang \Phi \rang$ and $\tilde{\Phi}$) does not contribute in the above expression, even when $\d\ln\eta/\d\psi$ is expressed in terms of \eqref{eq:eta}.

In \eqref{eq:fluxgS}--\eqref{eq:fluxqS}, the $|\nabla \psi|^2$ factors correspond to the $j_\perp$ factor in \eqref{eq:nctoc}, while the $j_\|$ factor in \eqref{eq:nctoc} arises due to the neoclassical transport \citep{braun2010a,helanderPRL2017}. In the following section, we will evaluate the above expression for example magnetic configurations. 

\section{Comparison to neoclassical calculations}
\label{sec:colscan}


In this section, we will compare the classical transport from \eqref{eq:fluxgS} to the neoclassical transport calculated with the drift-kinetic solver \textsc{Sfincs}.
Unlike analytical calculations of the neoclassical transport \citep{buller2018jpp,calvo2018nf}, this procedure is not limited to a specific collisionality regime, which will let us assess the importance of classical transport for any collisionality.

For this study, we will look at two stellarator configurations, where the neoclassical transport coefficients have been calculated across a wide range of collisionalities. Specifically, we will look at a simulated W7-X standard configuration case at the radial location $r_N=0.88$, with $T=\unit[1]{keV}$ and impurity parameters $Z=6$, $Z_{\text{eff}}=2.0$, studied by \citet{mollen2015}. The normalized radius is defined as $r_N = \sqrt{\psi_\text{t}/\psi_{\text{t},\text{LCFS}}}$, with $\psi_t$ the toroidal flux and $\psi_{\text{t},\text{LCFS}}$ its value at the last-closed flux-surface.
Since W7-X has been optimized for a low parallel current, and the standard configuration has low neoclassical transport compared to other configurations, we here expect the classical transport to dominate at high collisionality, as indicated at the end of \autoref{sec:motivation}.
In addition, we will look at a scenario based on an impurity hole plasma ($\#113208$, $t=4.64\,\rm{s}$, $r_N=0.6$, $T=\unit[3.2]{keV}$) of the Large Helical Device (LHD), where we replaced the mixture of helium and carbon impurities with purely carbon ($Z_\text{eff}=3.44$) for the sake of comparison. This magnetic configuration has been investigated in several studies, using both neoclassical \citep{velasco2017,mollen2018} and turbulence codes \citep{nunamiIAEA2016}. 

Effects of $\tilde{\Phi}$ and the radial electric field are not included in this demonstration (they are zero in the simulations), as this would make the drift-kinetic equation nonlinear, and add the complexity of finding the ambipolar electric field at each step.
These effects are not expected to strongly affect the classical transport, which is independent of the radial electric field, and typically not as sensitive to $\tilde{\Phi}$ as the neoclassical transport \citep{buller2018jpp}. The neoclassical transport can be both enhanced or reduced by these effects, which thus would affect the relative importance of classical transport. These effects will be touched upon in \autoref{sec:disc}.
As $\tilde{\Phi} = 0$ in this section, the density is a flux-function, and $\eta_a = n_a$. 

We scan the collisionality by artificially scaling the collision frequency.
For each point in the collisionality scan, we calculate the neoclassical and classical transport coefficients of the hydrogen bulk ion and the carbon impurity.
The transport coefficients for the (neo)classical fluxes are defined such that
\begin{equation}
  \Gamma^{\text{(N)C}}_{a}  = -n_{a} \left(D_{a,ni}^{\text{(N)C}} \frac{\d \ln n_i}{\d \psi} + D_{a,nz}^{\text{(N)C}} \frac{\d \ln n_z}{\d \psi} + D_{a,T}^{\text{(N)C}} \frac{\d \ln T}{\d \psi}
  \right), \label{eq:Ddef}
\end{equation}
where $a=i,z$ for ions and impurities. We have neglected the effects of electron collisions on the ion fluxes due to the small electron-to-ion mass-ratio, and assumed that the bulk and impurity ions have the same temperature $T\equiv T_i=T_z$. 

\begin{figure}
  \begin{subfigure}[b]{0.45\textwidth}
    \includegraphics[width=1.0\textwidth]{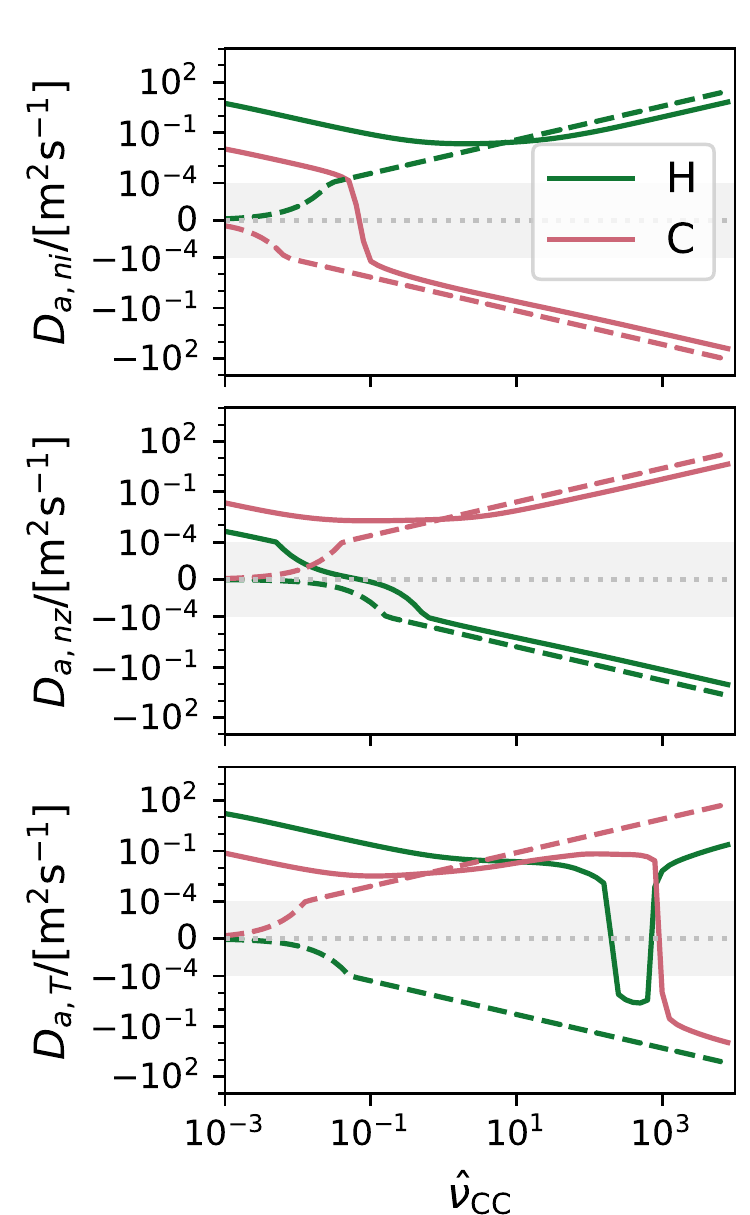}
  \end{subfigure}
  \begin{subfigure}[b]{0.45\textwidth}
    \includegraphics[width=1.0\textwidth]{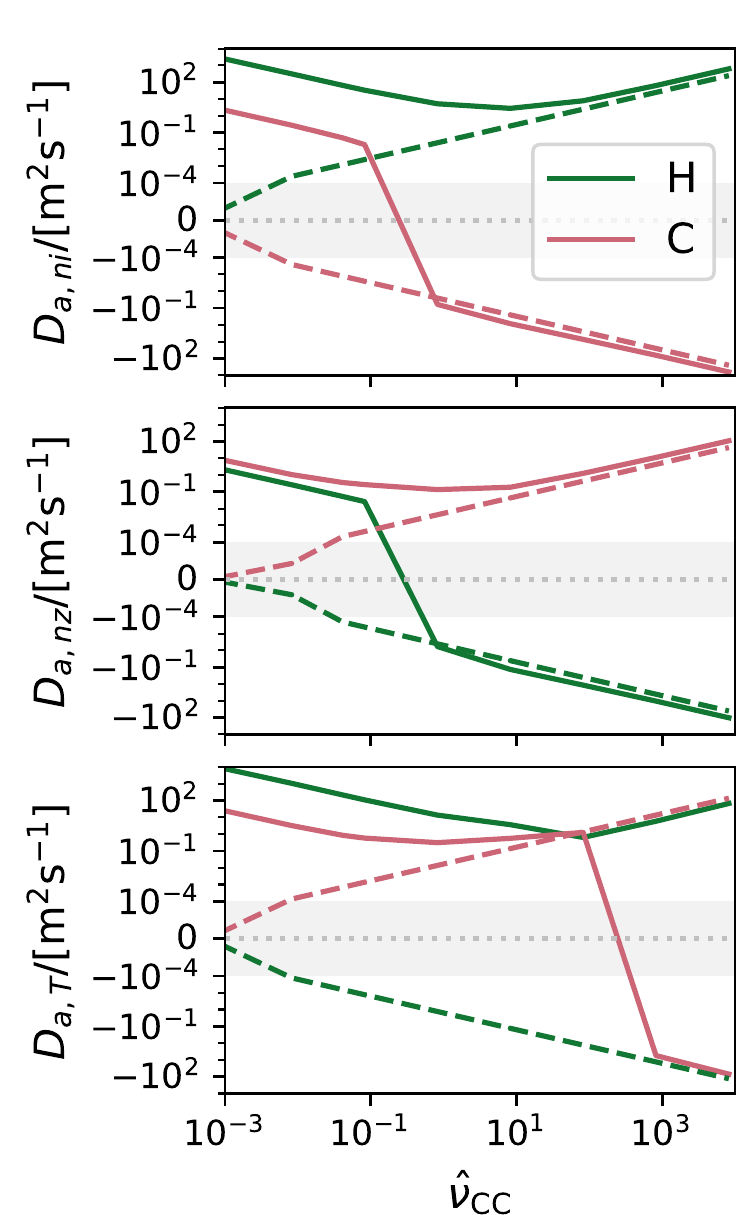}
  \end{subfigure}
  \caption{\label{fig:D} The neoclassical (\tikzsolid{black}) and classical (\tikzdashed{black}) transport coefficients as defined in \eqref{eq:Ddef}, plotted against the impurity-impurity collisionality. Left column: W7-X standard case. Right column: LHD impurity-hole case. The classical coefficients were calculated using \eqref{eq:fluxgS}, while the neoclassical coefficients were calculated using \textsc{Sfincs}. Note the \emph{symmetric logarithmic} scale of the y-axis; the shaded region has a linear y-axis scale.
}
\end{figure}

The results of the collisionality scan are shown in \autoref{fig:D}, with the collisionality defined as
\begin{equation}
\hat{\nu}_{ab} = \frac{G+\iota I}{B_{00} \sqrt{2T_a/m_a}} \frac{1}{\tau_{ab}},
\end{equation}
where $B_{00}$, $G$ and $I$ are related to the Boozer representation of the magnetic field (see, for example, \citet{mollen2018}) and $\iota$ is the rotational transform.
As seen in the left panels of \autoref{fig:D}, the impurity transport coefficients in the W7-X geometry are dominantly classical already for $\hat{\nu}_{CC} \gtrsim 1$. The cross-species contributions are dominantly classical already for $\hat{\nu}_{CC} \gtrsim 0.1$, for both the bulk and the impurity ions.
On the other hand, in LHD -- which has not been optimized for low $|j_\||/|\vec{j}_\perp|$ -- the classical transport for both species at most collisionalities is at least an order of magnitude smaller than the neoclassical transport.
An exception to this is the $D_{a,T}$ coefficient, where the classical transport becomes comparable or greater than the neoclassical transport at very high collisionalities ($\hat{\nu}_{CC} \gtrsim 100$). Another exception occurs in the collisionality range $\nu_{CC} \sim [0.1,1]$,  where the cross-species neoclassical $D_{z,ni}$ and $D_{i,nz}$ coefficients transition between different signs.

\section{Discussion}
\label{sec:disc}



\begin{figure}
  \begin{subfigure}[b]{0.45\textwidth}
    \includegraphics[width=1.0\textwidth]{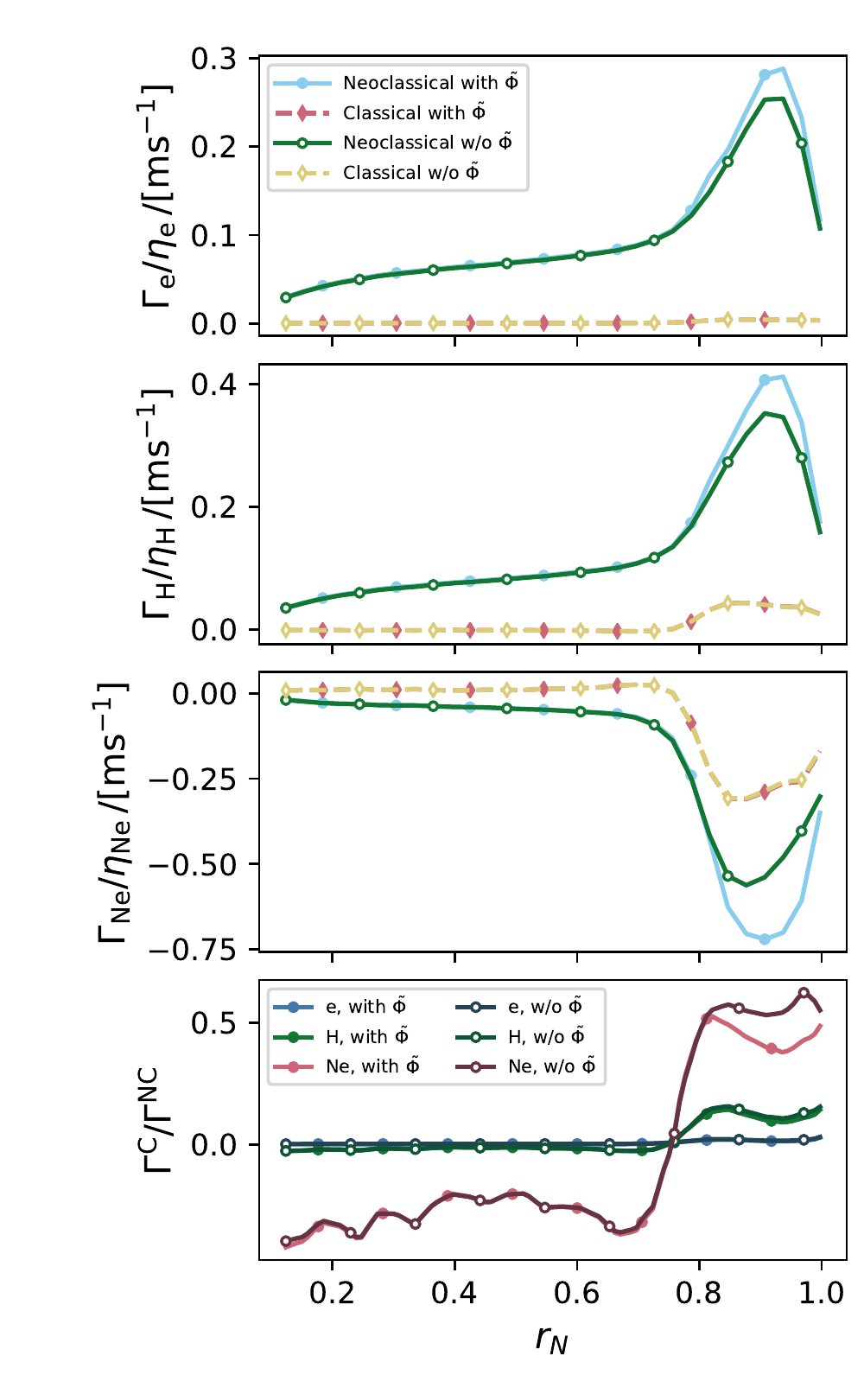}
  \end{subfigure}
  \begin{subfigure}[b]{0.45\textwidth}
    \includegraphics[width=1.0\textwidth]{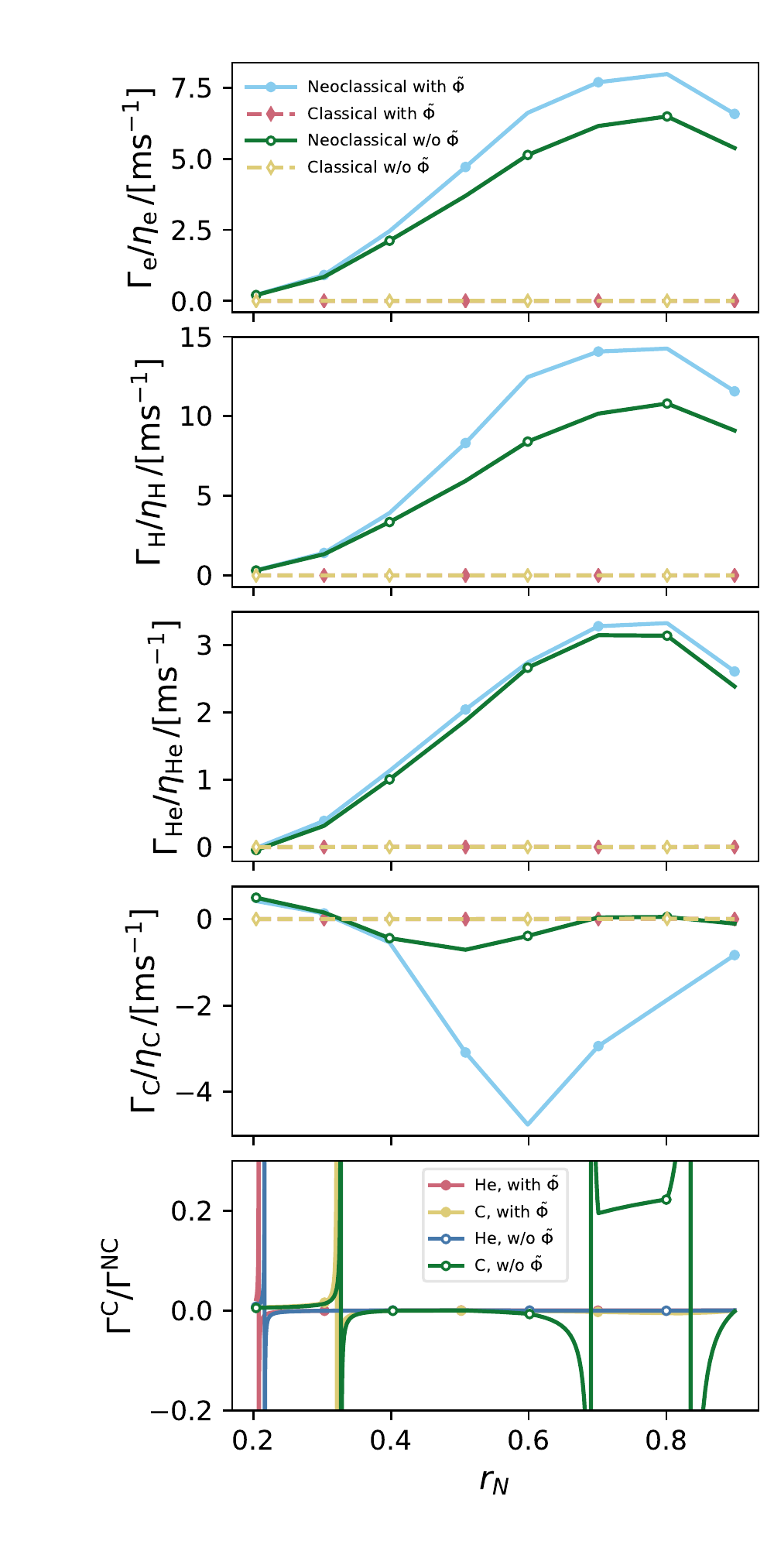}
  \end{subfigure}
  \caption{\label{fig:w7xLHD} Neoclassical (\tikzsolid{black}) and classical (\tikzdashed{black}) fluxes normalized to pseudo-densities for different species in W7-X (left) and LHD (right) as functions of the normalized radius $r_N$. Filled (open) symbols show the flux with (without) the effect of $\tilde{\Phi}$ included.
    The lowest panels show the ratio of the classical and neoclassical transport; note that this quantity diverges at radii where the neoclassical flux crosses zero.}
\end{figure}

We have seen that the neoclassical and classical transport coefficients can be comparable in a W7-X standard configuration, even at modest impurity collisionality ($\hat{\nu}_{\text{CC}} \gtrsim 0.1-1$), although the previously discussed simulations do not include effects of electrostatic potential variation within and across flux-surfaces and collisions with electrons. 
To demonstrate the relative importance of neoclassical and classical transport in realistic scenarios with these effects included, we consider two cases in which the full neoclassical behaviour has previously been analyzed by \citet{mollen2018}: a simulated neutral-beam heated high-mirror W7-X scenario with inward electric field and neon impurities, and the impurity hole LHD case of the previous section with an additional helium impurity. 

The classical and neoclassical fluxes are shown in \autoref{fig:w7xLHD} (left panels, W7X; right panels, LHD).
We note that in the W7-X case, the classical to neoclassical neon flux ratio is around $0.5$ in magnitude at most radii, and its sensitivity to a finite $\tilde{\Phi}$ is weak. This is consistent with neon being the only collisional impurity in this discharge (with $\hat{\nu}_{\text{NeNe}} \sim [0.5,2.5]$)

In the LHD case, the classical flux is generally small, although the classical to neoclassical flux ratio for carbon and helium diverges at discrete points, where the neoclassical flux crosses zero.
Additionally, for $\tilde{\Phi}=0$, there is a radial range ($0.6 \lesssim r_N \lesssim 0.8$), with small outward neoclassical carbon flux, where the neoclassical flux is only 5 times as large as the classical. 
When the effect of $\tilde{\Phi}$ is included, there is a large increase in the neoclassical carbon flux for $r_N \gtrsim 0.4$, with the result that the classical-to-neoclassical flux ratio remains small for these radii.


As all species in the LHD case are in a low collisionality regime, it is unlikely that the low neoclassical transport for $\tilde{\Phi}=0$ is due to small neoclassical transport coefficients, when compared to the classical coefficients. Rather, it may be that the contributions from the different neoclassical transport coefficients cancel out approximately. Including potential variations both changes the individual transport coefficient and somewhat reduces the ambipolar radial electric field in this case, both of which could affect this cancellation.

The classical fluxes are comparable to the neoclassical fluxes in W7-X, and should not generally be neglected in an analysis of the collisional transport.
Based on this conclusion, we have implemented the classical fluxes \eqref{eq:fluxgS}--\eqref{eq:fluxqS} as a post-processing step to the neoclassical codes \textsc{Sfincs} and \textsc{Dkes}; see the supplementary material for an example implementation in python.

As a final remark, we note that since the neoclassical transport in W7-X is sufficiently low to be comparable to the classical, the transport due to micro-turbulence can become relatively more important.
It may thus be necessary to consider the effect of turbulence on stellarator impurity transport in the future, which is often excluded due to the computational expense of simulating turbulence in stellarator geometry \citep{nunami2013}. Recent experimental studies by \citet{langenberg2018} and \citet{bgeiger2019} already point strongly in that direction. 

\acknowledgments SB and IP were supported by the International Career Grant of Vetenskapsr{\aa}det (Dnr.~330-2014-6313) and IP by Marie Sklodowska Curie Actions, Cofund, Project INCA 600398. SB's visit to Greifswald was supported by Chalmersska forskningsfonden.
This work has been carried out within the framework of the EUROfusion Consortium and has received funding from the Euratom research and training programme 2014-2018 and 2019-2020 under grant agreement No 633053. The views and opinions expressed herein do not necessarily reflect those of the European Commission.
The authors would like to thank the LHD experiment group and the technical staff of LHD for their support of this work. The authors are grateful to M.~Nunami for help with accessing the LHD data.



\appendix

\section{Braginskii matrix elements}
\label{sec:bme}
The Braginskii matrix elements are defined by
\begin{align}
  M_{ab}^{jk} = \frac{\tau_{ab}}{n_a} \int v_2 L_j^{(3/2)}(x_a^2) C_{ab}\left[\frac{m_a v_2}{T_a} L_k^{(3/2)}(x_a^2) f_{a0},f_{b0}\right] \d^3 v, \\
  N_{ab}^{jk} = \frac{\tau_{ab}}{n_a} \int v_2 L_j^{(3/2)}(x_a^2) C_{ab}\left[f_{a0},\frac{m_b v_2}{T_b} L_k^{(3/2)}(x_b^2)  f_{b0}\right] \d^3 v,
\end{align}
where $v_2$ is any Cartesian velocity component, $f_{a0}$ is a Maxwellian, $x_a = v/\sqrt{2T_a/m_a}$, $L_k^{(3/2)}$ are Sonine polynomials, where the polynomials relevant to classical particle and heat transport are
\begin{align}
  L_0^{(3/2)}(x_a^2) = 1, \\
  L_1^{(3/2)}(x_a^2) = \frac{5}{2} - x_a^2.
\end{align}
The corresponding relevant matrix elements are
\begin{align}
  M_{ab}^{00} =&  - \frac{\left(1+\frac{m_a}{m_b}\right)\left(1+\frac{m_a T_b}{m_b T_a}\right)}{\left(1+\frac{m_a T_b}{m_b T_a}\right)^{5/2}},\\
  M_{ab}^{01} =& - \frac{3}{2} \frac{1+\frac{m_a}{m_b}}{\left(1+\frac{m_a T_b}{m_b T_a}\right)^{5/2}}, \\
  M_{ab}^{11} =& - \frac{\frac{13}{4} + 4 \frac{m_a T_b}{m_b T_a} + \frac{15}{2}\left(\frac{m_a T_b}{m_b T_a}\right)^2 }{\left(1+\frac{m_a T_b}{m_b T_a}\right)^{5/2}}, \\
  N_{ab}^{11} =& \frac{27}{4}  \frac{\frac{m_a}{m_b}}{\left(1+\frac{m_a T_b}{m_b T_a}\right)^{5/2}}. \\
\end{align}

\bibliographystyle{jpp}
\bibliography{../../../plasma-bib} 

\begin{thebibliography}{23}
\expandafter\ifx\csname natexlab\endcsname\relax\def\natexlab#1{#1}\fi
\def\au#1{#1} \def\ed#1{#1} \def\yr#1{#1}\def\at#1{#1}\def\jt#1{\textit{#1}}
  \def\bt#1{#1}\def\bvol#1{\textbf{#1}} \def\vol#1{#1} \def\pg#1{#1}
  \def\publ#1{#1}\def\arxiv#1{#1}\def\org#1{#1}\def\st#1{\textit{#1}}

\bibitem[Beidler {\em et~al.\/}(2012)Beidler, Brakel, Burhenn, Dinklage,
  Erckmann, Feng, Geiger, Hartmann, Hirsch, Jaenicke, Koenig, Laqua,
  Maa{\ss}berg, Wagner, Weller \& Wobig]{fp2012iaea}
{\sc \au{Beidler, C.}, \au{Brakel, R.}, \au{Burhenn, R.}, \au{Dinklage, A.},
  \au{Erckmann, V.}, \au{Feng, Y.}, \au{Geiger, J.}, \au{Hartmann, D.},
  \au{Hirsch, M.}, \au{Jaenicke, R.}, \au{Koenig, R.}, \au{Laqua, H.},
  \au{Maa{\ss}berg, H.}, \au{Wagner, F.}, \au{Weller, A.} \& \au{Wobig, H.}}
  \yr{2012}  \bt{ \at{Fusion physics}}. chap. 8.3.6.  \publ{Vienna:
  International Atomic Energy Agency}.

\bibitem[Bolt {\em et~al.\/}(2002)Bolt, Barabash, Federici, Linke, Loarte, Roth
  \& Sato]{bolt2002}
{\sc \au{Bolt, H.}, \au{Barabash, V.}, \au{Federici, G.}, \au{Linke, J.},
  \au{Loarte, A.}, \au{Roth, J.} \& \au{Sato, K.}} \yr{2002}  \at{Plasma facing
  and high heat flux materials – needs for {ITER} and beyond}.  \jt{Journal
  of Nuclear Materials}  \bvol{307-311},  \pg{43 -- 52}.

\bibitem[Braginskii(1958)]{braginskii1958}
{\sc \au{Braginskii, S.}} \yr{1958}  \at{Transport phenomena in a completely
  ionized two-temperature plasma}.  \jt{Sov. Phys. JETP}  \bvol{6}~(33),
  \pg{358--369}.

\bibitem[Braun \& Helander(2010)]{braun2010a}
{\sc \au{Braun, S.} \& \au{Helander, P.}} \yr{2010}  \at{Pfirsch-{S}chl\"uter
  impurity transport in stellarators}.  \jt{Physics of Plasmas}  \bvol{17}~(7),
   \pg{072514}.

\bibitem[Buller {\em et~al.\/}(2018)Buller, Smith, Helander, Moll\'en, Newton
  \& Pusztai]{buller2018jpp}
{\sc \au{Buller, S.}, \au{Smith, H.~M.}, \au{Helander, P.}, \au{Moll\'en, A.},
  \au{Newton, S.~L.} \& \au{Pusztai, I.}} \yr{2018}  \at{Collisional transport
  of impurities with flux-surface varying density in stellarators}.
  \jt{Journal of Plasma Physics}  \bvol{84}~(4),  \pg{905840409}.

\bibitem[Calvo {\em et~al.\/}(2018)Calvo, Parra, Velasco, Alonso \&
  {n}a]{calvo2018nf}
{\sc \au{Calvo, I.}, \au{Parra, F.~I.}, \au{Velasco, J.~L.}, \au{Alonso, J.~A.}
  \& \au{{n}a, J. G.-R.}} \yr{2018}  \at{Stellarator impurity flux driven by
  electric fields tangent to magnetic surfaces}.  \jt{Nuclear Fusion}
  \bvol{58}~(12),  \pg{124005}.

\bibitem[Garc\'{i}a-Rega{\~{n}}a {\em et~al.\/}(2017)Garc\'{i}a-Rega{\~{n}}a,
  Beidler, Kleiber, Helander, Moll\'{e}n, Alonso, Landreman, Maa{\ss}berg,
  Smith, Turkin \& Velasco]{garcia2017}
{\sc \au{Garc\'{i}a-Rega{\~{n}}a, J.}, \au{Beidler, C.}, \au{Kleiber, R.},
  \au{Helander, P.}, \au{Moll\'{e}n, A.}, \au{Alonso, J.}, \au{Landreman, M.},
  \au{Maa{\ss}berg, H.}, \au{Smith, H.}, \au{Turkin, Y.} \& \au{Velasco, J.}}
  \yr{2017}  \at{Electrostatic potential variation on the flux surface and its
  impact on impurity transport}.  \jt{Nuclear Fusion}  \bvol{57}~(5),
  \pg{056004}.

\bibitem[Geiger {\em et~al.\/}(2019)Geiger, Wegner, Beidler, Burhenn,
  Buttensch\"on, Dux, Langenberg, Pablant, P\"utterich, Turkin, Windisch,
  Winters, Beurskens, Biedermann, Brunner, Cseh, Damm, Effenberg, Fuchert,
  Grulke, Harris, Killer, Knauer, Kocsis, Kr\"amer-Flecken, Kremeyer,
  Krychowiak, Marchuk, Nicolai, Rahbarnia, Satheeswaran, Schilling, Schmitz,
  Schr\"oder, Szepesi, Thomsen, Mora, Traverso, Zhang \& {The W7-X
  Team}]{bgeiger2019}
{\sc \au{Geiger, B.}, \au{Wegner, T.}, \au{Beidler, C.}, \au{Burhenn, R.},
  \au{Buttensch\"on, B.}, \au{Dux, R.}, \au{Langenberg, A.}, \au{Pablant, N.},
  \au{P\"utterich, T.}, \au{Turkin, Y.}, \au{Windisch, T.}, \au{Winters, V.},
  \au{Beurskens, M.}, \au{Biedermann, C.}, \au{Brunner, K.}, \au{Cseh, G.},
  \au{Damm, H.}, \au{Effenberg, F.}, \au{Fuchert, G.}, \au{Grulke, O.},
  \au{Harris, J.}, \au{Killer, C.}, \au{Knauer, J.}, \au{Kocsis, G.},
  \au{Kr\"amer-Flecken, A.}, \au{Kremeyer, T.}, \au{Krychowiak, M.},
  \au{Marchuk, O.}, \au{Nicolai, D.}, \au{Rahbarnia, K.}, \au{Satheeswaran,
  G.}, \au{Schilling, J.}, \au{Schmitz, O.}, \au{Schr\"oder, T.}, \au{Szepesi,
  T.}, \au{Thomsen, H.}, \au{Mora, H.~T.}, \au{Traverso, P.}, \au{Zhang, D.} \&
  \au{{The W7-X Team}}} \yr{2019}  \at{Observation of anomalous impurity
  transport during low-density experiments in {W7-X} with laser blow-off
  injections of iron}.  \jt{Nuclear Fusion}  \bvol{59}~(4),  \pg{046009}.

\bibitem[{Hazeltine}(1973)]{hazeltine1973}
{\sc \au{{Hazeltine}, R.~D.}} \yr{1973}  \at{Recursive derivation of
  drift-kinetic equation}.  \jt{Plasma Physics}  \bvol{15}~(1),  \pg{77}.

\bibitem[Helander {\em et~al.\/}(2017)Helander, Newton, Moll\'en \&
  Smith]{helanderPRL2017}
{\sc \au{Helander, P.}, \au{Newton, S.~L.}, \au{Moll\'en, A.} \& \au{Smith,
  H.~M.}} \yr{2017}  \at{Impurity transport in a mixed-collisionality
  stellarator plasma}.  \jt{Phys. Rev. Lett.}  \bvol{118},  \pg{155002}.

\bibitem[Helander \& Sigmar(2005)]{helander2005}
{\sc \au{Helander, P.} \& \au{Sigmar, D.~J.}} \yr{2005} {\em Collisional
  Transport in Magnetized Plasmas\/}.  \publ{Cambridge University Press}.

\bibitem[Hinton \& Wong(1985)]{hintonWong1985}
{\sc \au{Hinton, F.~L.} \& \au{Wong, S.~K.}} \yr{1985}  \at{Neoclassical ion
  transport in rotating axisymmetric plasmas}.  \jt{The Physics of Fluids}
  \bvol{28}~(10),  \pg{3082--3098}.

\bibitem[Landreman {\em et~al.\/}(2014)Landreman, Smith, Moll\'{e}n \&
  Helander]{sfincs2014}
{\sc \au{Landreman, M.}, \au{Smith, H.~M.}, \au{Moll\'{e}n, A.} \&
  \au{Helander, P.}} \yr{2014}  \at{Comparison of particle trajectories and
  collision operators for collisional transport in nonaxisymmetric plasmas}.
  \jt{Physics of Plasmas}  \bvol{21}~(4),  \pg{042503}.

\bibitem[Langenberg {\em et~al.\/}(2018)Langenberg, Warmer, Fuchert, Marchuk,
  Dinklage, Wegner, Alonso, Bozhenkov, Brunner, Burhenn, Buttenschön, Drews,
  Geiger, Grulke, Hirsch, H\"ofel, Hollfeld, Killer, Knauer, Krings, Kunkel,
  Neuner, Offermanns, Pablant, Pasch, Rahbarnia, Satheeswaran, Schilling,
  Schweer, Thomsen, Traverso, Wolf \& {the W7-X Team}]{langenberg2018}
{\sc \au{Langenberg, A.}, \au{Warmer, F.}, \au{Fuchert, G.}, \au{Marchuk, O.},
  \au{Dinklage, A.}, \au{Wegner, T.}, \au{Alonso, J.~A.}, \au{Bozhenkov, S.},
  \au{Brunner, K.~J.}, \au{Burhenn, R.}, \au{Buttenschön, B.}, \au{Drews, P.},
  \au{Geiger, B.}, \au{Grulke, O.}, \au{Hirsch, M.}, \au{H\"ofel, U.},
  \au{Hollfeld, K.~P.}, \au{Killer, C.}, \au{Knauer, J.}, \au{Krings, T.},
  \au{Kunkel, F.}, \au{Neuner, U.}, \au{Offermanns, G.}, \au{Pablant, N.~A.},
  \au{Pasch, E.}, \au{Rahbarnia, K.}, \au{Satheeswaran, G.}, \au{Schilling,
  J.}, \au{Schweer, B.}, \au{Thomsen, H.}, \au{Traverso, P.}, \au{Wolf, R.~C.}
  \& \au{{the W7-X Team}}} \yr{2018}  \at{Impurity transport studies at
  {Wendelstein 7-X} by means of x-ray imaging spectrometer measurements}.
  \jt{Plasma Physics and Controlled Fusion}  \bvol{61}~(1),  \pg{014030}.

\bibitem[Moll\'{e}n {\em et~al.\/}(2015)Moll\'{e}n, Landreman, Smith, Braun \&
  Helander]{mollen2015}
{\sc \au{Moll\'{e}n, A.}, \au{Landreman, M.}, \au{Smith, H.~M.}, \au{Braun, S.}
  \& \au{Helander, P.}} \yr{2015}  \at{Impurities in a non-axisymmetric plasma:
  Transport and effect on bootstrap current}.  \jt{Physics of Plasmas}
  \bvol{22}~(11),  \pg{112508}.

\bibitem[Moll\'en {\em et~al.\/}(2018)Moll\'en, Landreman, Smith,
  {Garc\'ia-Rega\~{n}a} \& Nunami]{mollen2018}
{\sc \au{Moll\'en, A.}, \au{Landreman, M.}, \au{Smith, H.~M.},
  \au{{Garc\'ia-Rega\~{n}a}, J.~M.} \& \au{Nunami, M.}} \yr{2018}
  \at{Flux-surface variations of the electrostatic potential in stellarators:
  impact on the radial electric field and neoclassical impurity transport}.
  \jt{Plasma Physics and Controlled Fusion}  \bvol{60}~(8),  \pg{084001}.

\bibitem[Newton \& Helander(2006)]{newton2006}
{\sc \au{Newton, S.} \& \au{Helander, P.}} \yr{2006}  \at{Neoclassical momentum
  transport in an impure rotating tokamak plasma}.  \jt{Physics of Plasmas}
  \bvol{13}~(1),  \pg{012505}.

\bibitem[Newton {\em et~al.\/}(2017)Newton, Helander, Moll\'{e}n \&
  Smith]{newton2017}
{\sc \au{Newton, S.~L.}, \au{Helander, P.}, \au{Moll\'{e}n, A.} \& \au{Smith,
  H.~M.}} \yr{2017}  \at{Impurity transport and bulk ion flow in a mixed
  collisionality stellarator plasma}.  \jt{Journal of Plasma Physics}
  \bvol{83}~(5),  \pg{905830505}.

\bibitem[N\"uhrenberg \& Zille(1986)]{nuhrenberg1986}
{\sc \au{N\"uhrenberg, J.} \& \au{Zille, R.}} \yr{1986}  \at{Stable
  stellarators with medium $\beta$ and aspect ratio}.  \jt{Physics Letters A}
  \bvol{114}~(3),  \pg{129 -- 132}.

\bibitem[Nunami {\em et~al.\/}(2016)Nunami, Sugama, Velasco, Yokoyama, Sato,
  Nakata \& Satake]{nunamiIAEA2016}
{\sc \au{Nunami, M.}, \au{Sugama, H.}, \au{Velasco, J.~L.}, \au{Yokoyama, M.},
  \au{Sato, M.}, \au{Nakata, M.} \& \au{Satake, S.}} \yr{2016}  \at{Anomalous
  and neoclassical transport of hydrogen isotope and impurity ions in {LHD}
  plasmas}.  \jt{IAEA Fusion Energy Conference preprints, Kyoto, Japan}
  \pg{pp. TH/P2--3}.

\bibitem[Nunami {\em et~al.\/}(2013)Nunami, Watanabe \& Sugama]{nunami2013}
{\sc \au{Nunami, M.}, \au{Watanabe, T.-H.} \& \au{Sugama, H.}} \yr{2013}  \at{A
  reduced model for ion temperature gradient turbulent transport in helical
  plasmas}.  \jt{Physics of Plasmas}  \bvol{20}~(9),  \pg{092307}.

\bibitem[Pfirsch \& Schl{\"u}ter(1962)]{ps1962}
{\sc \au{Pfirsch, D.} \& \au{Schl{\"u}ter, A.}} \yr{1962}  \at{Der einfluss der
  elektrischen leitf{\"a}higkeit auf das gleichgewichtsverhalten von plasmen
  niedrigen drucks in stelleratoren}.  \jt{Max-Planck-Institut Report
  MPI/PA/7/62}  \pg{pp. 88--89}.

\bibitem[Velasco {\em et~al.\/}(2017)Velasco, Calvo, Satake, Alonso, Nunami,
  Yokoyama, Sato, Estrada, Fontdecaba, Liniers, McCarthy, Medina, Milligen,
  Ochando, Parra, Sugama, Zhezhera, {The LHD Experimental Team} \& {The TJ-II
  Team}]{velasco2017}
{\sc \au{Velasco, J.}, \au{Calvo, I.}, \au{Satake, S.}, \au{Alonso, A.},
  \au{Nunami, M.}, \au{Yokoyama, M.}, \au{Sato, M.}, \au{Estrada, T.},
  \au{Fontdecaba, J.}, \au{Liniers, M.}, \au{McCarthy, K.}, \au{Medina, F.},
  \au{Milligen, B. P.~V.}, \au{Ochando, M.}, \au{Parra, F.}, \au{Sugama, H.},
  \au{Zhezhera, A.}, \au{{The LHD Experimental Team}} \& \au{{The TJ-II Team}}}
  \yr{2017}  \at{Moderation of neoclassical impurity accumulation in high
  temperature plasmas of helical devices}.  \jt{Nuclear Fusion}  \bvol{57}~(1),
   \pg{016016}.

\end{thebibliography}

\end{document}